\documentclass{mscs}

\usepackage{amsmath}
\usepackage{amssymb}
\usepackage{stmaryrd}
\usepackage[english]{babel}
\usepackage{euscript}
\usepackage{graphicx}
\usepackage{picins}
\usepackage{supertabular}
\usepackage{float}
\floatstyle{ruled}
\newfloat{programme}{htbp}{prog}
\floatname{programme}{{\bf Program}}
\usepackage{color}
\definecolor{darkred}{rgb}{0.90,0.05,0.04}

\newcommand{\syntaxdef}{::=}
\newcommand{\procdef}{\displaystyle \mathop{=}^{\mathtt{def}}}
\newcommand{\nattype}{\mathtt{Nat}}
\newcommand{\qubittype}{\mathtt{Qubit}}

\newcommand{\qvar}[1]{#1:\qubittype}
\newcommand{\cvar}[1]{#1:\nattype}
\newcommand{\declparam}[1]{\pmb [\ #1\ \pmb] \bullet}
\newcommand{\declvar}[2]{#1: #2}
\newcommand{\findeclvar}{\pmb\ .\ }
\newcommand{\debutbloc}{\pmb [\ }
\newcommand{\finblocse}{\pmb ]}
\newcommand{\finbloc}{\ \pmb ]}
\newcommand{\prefix}{.}
\newcommand{\seq}{\ ;}
\newcommand{\para}{\parallel}
\newcommand{\probbin}[1]{\oplus_{#1}}
\newcommand{\prob}[1]{\bigoplus_{#1}}
\newcommand{\nondet}{+}
\newcommand{\cond}[2]{#1 \rightarrow #2}
\newcommand{\restrict}[1]{\backslash#1\ }
\newcommand{\restrictg}[1]{\backslash\{#1\}\ }
\newcommand{\envoi}[1]{\ !#1\  }
\newcommand{\recep}[1]{\ ?#1\  }
\newcommand{\term}{\mbox{\it end}}
\newcommand{\stopproc}{\mbox{\it nil}}
\newcommand{\comment}[1]{\mbox{// #1}}
\newcommand{\superlongrightarrow}{-\!-\hspace{-5pt}\longrightarrow}
\newcommand{\supermathoparrow}{\mathop{\superlongrightarrow}}
\newcommand{\mathoparrow}{\mathop{\longrightarrow}}

\newcommand{\atransition}[1]{\mbox{$\displaystyle\ \supermathoparrow^{#1}\ $}}
\newcommand{\petiteatransition}{\xrightarrow{\ a\ }}
\newcommand{\tautransition}{\mbox{$\displaystyle\ \mathoparrow^{\tau}\ $}}

\newcommand{\ptransition}[1]{\longrightarrow_{#1}}
\newcommand{\divtransition}{\dashrightarrow}
\newcommand{\vtransition}{\longrightarrow_v}
\newcommand{\etransition}{\longrightarrow_e}
\newcommand{\ket}[1]{|#1\rangle}
\newcommand{\bra}[1]{\langle#1|}
\newcommand{\contexte}[4]{<#1, #2 = #3, #4 >}

\newcommand{\spcontexte}[1]{/ #1}
\newcommand{\contexteqcq}{<s, q = \rho, f >}
\newcommand{\scontexteqcq}{/<s, q = \rho, f >}
\newcommand{\symbcontexteprob}{\boxplus}
\newcommand{\bigsymbcontexteprob}{\mathop\boxplus}
\newcommand{\contexteprob}[5]{\bigsymbcontexteprob_{#1}\!\!<#2, #3 = #4, #5 >}
\newcommand{\contexteprobqcq}{\bigsymbcontexteprob_{p_i}\!\!<s_i, q_i = \rho_i, f_i>}
\newcommand{\pcontexteprob}[2]{\bigsymbcontexteprob_{#1}#2}

\newcommand{\spcontexteprob}[2]{/\bigsymbcontexteprob_{#1}#2}
\newcommand{\spcontexteprobbin}[3]{/ #2\symbcontexteprob_{#1}#3}
\newcommand{\contextestable}{\downarrow}
\newcommand{\stable}{\downarrow}
\newcommand{\trace}[1]{Tr(#1)}
\newcommand{\traceout}[3]{Tr_{#1/#2}(#3)}
\newcommand{\varpile}[1]{\mathtt{Var}(#1)} 
\newcommand{\pileconcat}{|} 
\newcommand{\pileajout}{.}
\newcommand{\pileajoutp}[2]{#2.#1}
\newcommand{\rmpile}[2]{#1\backslash#2}
\newcommand{\dom}[1]{\mbox{dom($#1$)}}

\newcommand{\tailleseq}[1]{\mathtt{size}(#1)}
\newcommand{\gate}[1]{\mbox{\it #1}}
\newcommand{\proc}[1]{\mbox{\bf #1}}

\newcommand{\enstransfadm}{\mathcal A}
\newcommand{\mstdp}[2]{M_{std, #1}[#2]}
\newcommand{\mstd}[1]{M_{std, #1}}

\newcommand{\adjoint}[1]{#1^\dagger}
\newcommand{\appli}[1]{\mathcal{T}_{#1}}
\newcommand{\applip}[2]{\appli{#1}(#2)}
\newcommand{\n}{I\!\! N}
\newcommand{\ensemble}[1]{\{#1\}}
\newcommand{\card}[1]{|#1|}
\newcommand{\norme}[1]{\|#1\|}
\newcommand{\abs}[1]{|#1|}
\newcommand{\petitsautdeligne}{\vspace{5pt}}
\newcommand{\sautdeligne}{\vspace{10pt}}
\newcommand{\tab}{\hspace{20pt}}
\newtheorem{definition}{Definition}
\newtheorem{proposition}{Proposition}
\newenvironment{preuve}{\par\noindent\textit{Proof.}\ }{\hfill$\diamond$\par}
\newenvironment{remarque}{\par\noindent\textit{Note.}\ }{\par}
\newcommand{\bisimul}{\mathcal{B}}
\newcommand{\ensetats}{\EuScript{S}}
\newcommand{\siltransition}{\leadsto}
\newcommand{\seqsiltransition}{\siltransition^*}
\newcommand{\bisimilar}{\leftrightarroweq}
\newcommand{\rootedbisimilar}{\leftrightarroweq_{r}}
\newcommand{\etatqvar}[2]{\rho_{#1}^{#2}}
\newcommand{\classeeq}[1]{\bar{#1}}
\newcommand{\parties}[1]{\mathcal{P}(#1)}
\newcommand{\atteint}{\triangleright}
\newcommand{\eqrel}{\mathcal{R}}

\newcommand{\congrusem}{\sim}
\newcommand{\eprinit}{\proc{BuildEPR}}

\newcommand{\alice}{\proc{Alice}}
\newcommand{\bob}{\proc{Bob}}
\newcommand{\teleport}{\proc{Teleport}}

\newcommand{\regleinf}[3]
{\begin{eqnarray}\label{#1}
\frac{#2}{#3}
\end{eqnarray}}
\newcommand{\regleinfcond}[4]
{\begin{eqnarray}\label{#1}
\frac{#2}{#3}\tab #4
\end{eqnarray}}

\title {Relations among Quantum Processes: Bisimilarity and Congruence}
\author[M. Lalire]{M\ls A\ls R\ls I\ls E\ns L\ls A\ls L\ls I\ls R\ls E\\
Leibniz Laboratory - Grenoble, France}

\date{25 January 2005; Revised 23 November 2005}
 
\begin{document}

\label{firstpage}
\maketitle

\begin{abstract}
Full formal descriptions of algorithms making use of quan\-tum principles must take into account both quan\-tum and classical computing components, as well as communications between these components.
Moreover, to model concurrent and distributed quan\-tum computations and quan\-tum communication protocols, communications over quantum channels which move qubits physically from one place to another must also be taken into account.

Inspired by classical process algebras, which provide a framework for modeling cooperating computations, a process algebraic notation is defined. This notation provides a homogeneous style to formal descriptions of concurrent and distributed computations comprising both quan\-tum and classical parts.
Based upon an operational semantics which makes sure that quantum objects, operations and 
communications operate according to the postulates of quantum mechanics, an equivalence is defined among process states considered as having the same behavior. This equivalence is a probabilistic branching bisimulation. From this relation, an equivalence on processes is defined. However, it is not a congruence because it is not preserved by parallel composition.
\end{abstract}

\section {Introduction}
\label{sectionIntroduction}

The number of quantum programming languages is growing rapidly. These languages can be classified in three families: imperative, functional, and parallel and distributed.
Among imperative programming languages, there are  QCL \cite{Omer00QCL}, designed by \"Omer, which aims at simulating quantum programs, and qGCL \cite{Zuliani01These} by Zuliani which allows the construction by refinement of proved correct quantum programs.
QPL \cite{Selinger04QPL} is a functional language designed by Selinger with a denotational semantics. Several quantum $\lambda$-calculi have also been developed: for example \cite{Tonder03LambdaCalculus} by Van Tonder, based on a simplified linear $\lambda$-calculus and \cite{ArrighiDowek04Turku} by Arrighi and Dowek, which is a "linear-algebraic $\lambda$-calculus".
Gay and Nagarajan have developed CQP, a language to describe communicating quantum processes \cite{NagGay04Turku}. This language is based on $\pi$-calculus. An important point in their work is the definition of a type system, and the proof that the operational semantics preserves typing.
  
Cooperation between quantum and classical computations is inherent in quan\-tum algorithmics.
Teleportation of a qubit state from Alice to Bob \cite{BennettBrassard93Teleportation} is a good example of this cooperation. Indeed, Alice carries out a measurement, the result of which (two bits) is sent to Bob, and Bob uses this classical result to determine which quantum transformation he must apply.
Moreover, initial preparation of quantum states and measurement of quantum results are two essential forms of interactions between the classical and quantum parts of computations which a language must be able to express.
Process algebras are a good candidate for such a language since they provide a framework for modeling cooperating computations.
In addition, they have well defined semantics and permit the transformation of programs as well as the formal study and analysis of their properties.
Their semantics give rise to an equivalence relation on processes. Bisimilarity is an adequate equivalence relation to deal with communicating processes since it relates processes that can execute the same flows of actions while having the same branching structure.

This paper presents first the main points of the definition and semantics of a Quantum Process Algebra (QPAlg). Then, a probabilistic branching bisimilarity is defined among process states (section \ref{sectionBisimul}), this relation is proved to be an equivalence. As an example, in section \ref{sectionHad}, the application of the Hadamard unitary transformation is proved bisimilar with its simulation with measurement only, based on state transfer. Finally, in section \ref{sectionCongru}, an equivalence relation among processes is defined. This relation is preserved by all the operators of QPAlg except parallel composition.

\section {Definition of QPAlg}
\label{sectionQuantProcAlg}

The process algebra QPAlg is based upon process algebras such as CCS \cite{Milner89CommConc} and Lotos \cite{Bolognesi87IsoLotos}.
The key aspects of QPAlg are developed in this section. The precise syntax and the main inference rules of the semantics are given in appendix \ref{annexQPA}. For more details and more examples, see \cite{LalireJorrand04Turku}.

\subsection {Variables}
\label{subsecQuantVar}

For the purpose of this paper, we consider two types of variables, one classical: {\it $\nattype$}, for variables taking natural values, and one quan\-tum: {\it $\qubittype$} for variables standing for qubits. An extended version of the process algebra would of course also include quantum registers and other types of variables.

In classical process algebras with value passing
\cite{Milner89CommConc, Bolognesi87IsoLotos, Roscoe98CSP},
variables are instantiated when communications between processes  occur and cannot be modified after their instantiation. As a consequence, it is not necessary to store their values.
In fact, when a variable is instantiated, all its occurrences are replaced by the value received.

Here, quan\-tum variables represent physical qubits. Applying a transformation to a variable which represents a qubit modifies the state of that qubit.
This means that values of variables are modified. For that reason, a process state must keep track of both variable names and variable states, this is achieved thanks to the context (cf. section \ref{subseccontext}).

Variables are declared using the following syntax:
$\debutbloc \declvar{x_1}{t_1},\ldots,\declvar{x_n}{t_n} \findeclvar P \finbloc$
 where  $x_1,\ldots,x_n$ is a list of variables, $t_1,\ldots,t_n$ are their types, and $P$ is a process which can make use of these classical and quan\-tum variables.

\subsection {Expressions}

The quantum expressions are quantum variables or tensor product of quantum variables, denoted $x_1 \otimes \cdots \otimes x_n$.

The classical expressions are usual classical expressions, and application of an admissible transformation to a quantum expression. Admissible transformations are also called general quantum measurements, it includes unitary transformations. For more details, see \cite{NielsenChuang00Book}.

Let $\enstransfadm$ be a set of predefined admissible transformations. The application of the admissible transformation $A=\ensemble{A_{\tau_1},\ldots,A_{\tau_m}} \in \enstransfadm$, of dimension $n$, to the register of qubits $x_1\otimes\cdots\otimes x_n$ is denoted $A[x_1\otimes\cdots\otimes x_n]$.

The quantum memory is stored in the context in the form $q = \rho$ where $q$ is the list of quantum variable names and $\rho$, a density matrix representing their quantum state (cf. section \ref{subseccontext}).
If the classical result of $A[x_1\otimes\cdots\otimes x_n]$ is $\tau_i$, then $\appli{A_{\tau_i}}$ is the super-operator which must be applied to the density matrix $\rho$, to describe the evolution of the quantum memory $q=\rho$.
$$\appli{A_{\tau_i}}: \rho \mapsto\adjoint{\Pi}.(A_{\tau_i}\otimes I^{\otimes k}).\Pi.\rho.\adjoint{\Pi}.
(\adjoint{A_{\tau_i}}\otimes I^{\otimes k}).\Pi$$
where
\begin{itemize}
\item $\Pi$ is the permutation matrix which places the $x_i$'s at the head of $q$
\item $k = \tailleseq q - n\ $
\item $I^{\otimes k}\!= \underbrace{I \otimes \cdots \otimes I}_k$, where $I$ is the identity matrix on $\mathbb C^2$
\end{itemize}

Since the admissible transformation $A_{\tau_i}$ may be applied to qubits which are anywhere within the list $q$, a permutation $\Pi$ must be applied first. This permutation moves the $x_i$'s so that they are placed at the head of $q$ in the order specified by $x_1,\ldots,x_n$.
Then $A_{\tau_i}$ can be applied to the first $n$ elements and $I$ to the remainder. Finally, the last operation is the inverse of the permutation $\Pi$
so that at the end, the arrangement of the elements in $\rho$ is consistent with the order of the elements in $q$.

$A[x_1\otimes\cdots\otimes x_n]$ is not probabilistic but its evaluation produces a result with a probabilistic value. This value is stored in the context which becomes probabilistic (cf. section \ref{subseccontext}).

In the examples of this paper, the set $\enstransfadm$ of admissible transformations is:
$$\enstransfadm = \{H, CNot, I, \sigma_X, \sigma_Y, \sigma_Z, \mstd{1}, \mstd{2}, X, Z\otimes X\}$$
$H$ is Hadamard transformation, $CNot$ is "controlled not", $I$ is the identity, and $\sigma_X$, $\sigma_Y$, $\sigma_Z$ are Pauli operators. $\mstd{1}$ and $\mstd{2}$ correspond to measurement in the standard basis of respectively one and two qubits. $X$ and $Z\otimes X$ are the admissible transformations corresponding to the measurements with the Pauli observables $X$ and $Z\otimes X$.

\subsection {Basic actions}

The basic actions of QPAlg are classical expressions and communications. A classical expression is interesting as a basic action if it has side effects, as it is the case of the application of an admissible transformation.

There are several kinds of communications, depending on the type of the expression sent and the type of the receiving variable. The different kinds of communications are: classical to classical communications, classical to quantum communications for initializing qubits, and quantum to quantum communications for allowing the description of quantum communication protocols.
Communication gates are not typed but we can imagine a subsequent version of QPAlg where communication gates would be declared with a fixed type like the other variables.

Emission of an expression $e$ from a gate $g$ is denoted $g \envoi e$, reception in a variable $x$ is denoted $g \recep x$.
In the operational semantics of parallel composition (rules \ref{infcomppara} to \ref{infcompparaprob2} of the semantics given in appendix \ref{annexQPASem}), the combination of the rules for emission and reception defines communication.
In a classical to quan\-tum communication (rule \ref{infcommcq}), the qubit is initialized in the basis state $\ket v \bra v$, where $v$ is the classical value sent (in this case, $v$ must be $0$ or $1$). In a quantum to quantum communication (rule \ref{infcommqq}), the name of the sent qubit is replaced in the context by the name of the receiving qubit.

\subsection {Composition operators}

To create a process from basic actions, the prefix operator "$\prefix$" is used: if $\alpha$ is an action and $P$, a process, $\alpha\prefix P$ is a new process which performs $\alpha$ first, then behaves as $P$.

The predefined process $\stopproc$ cannot perform any transition.

The operators of the process algebra are: parallel composition ($P\para Q$), nondeterministic choice ($P \nondet Q$), probabilistic choice ($P\probbin{p}Q$), conditional choice
($\debutbloc \cond{c_1}{P_1},\ldots,\cond{c_n}{P_n}\finbloc$) and restriction ($P\restrict L$).
The process $\debutbloc \cond{c_1}{P_1},\ldots,\cond{c_n}{P_n}\finbloc$, where $c_i$ is a condition and $P_i$ a process, evolves as a process chosen nondeterministically among the processes $P_j$ such that $c_j$ is true.
Restriction is useful for disallowing the use of some gates (the gates listed in $L$),
thus forcing internal communication within process $P$. 
Communication can occur between two parallel processes whenever a value emission in one of them and a value reception in the other one use the same gate name.

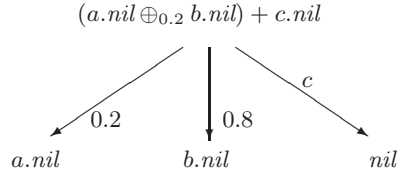
\begin{figure}[hbtp]
\begin{center}
\begin{picture}(160,75)
\put(70,50){\vector(-3,-2){50}}
\put(80,50){\vector(0,-1){35}}
\put(90,50){\vector(3,-2){50}}
\put(35,20){\mbox{$0.2$}}
\put(85,20){\mbox{$0.8$}}
\put(115,35){\mbox{$c$}}
\put(30,60){\mbox{$(a\prefix \stopproc \probbin{0.2} b\prefix \stopproc)
+ c\prefix \stopproc$}}
\put(5,5){\mbox{$a\prefix\stopproc$}}
\put(70,5){\mbox{$b\prefix\stopproc$}}
\put(140,5){\mbox{$\stopproc$}}
\end{picture}
\end{center}
\caption {Nondeterminism solved before probabilistic choice}
\label{figNondetProb}
\end{figure}

The process $P\probbin{p}Q$ behaves like $P$ with probability $p$ and like $Q$ with probability $1-p$. As explained in \cite{Cazorla01Art} and \cite{CazorlaMis} and shown in the example of figure \ref{figNondetProb}, if a process contains both a probabilistic and a nondeterministic choice, then the probabilistic choice must always be solved first. Otherwise, a probabilistic transition labeled with a probability $p$ does not mean that this transition will be executed with probability $p$.

To solve probabilistic choices before nondeterministic ones, the notion of stability for processes is defined.
\begin{definition}
\textbf {Probabilistic stability} is defined by induction:
\begin{enumerate}
\item $\stopproc$, $a\prefix P$,
$\debutbloc \declvar{x_1}{t_1},\ldots \declvar{x_n}{t_n} \findeclvar P\finbloc$ are stable.
\item $P\restrict L$, $\declparam{\declvar{x_1}{t_1},\ldots, \declvar{x_n}{t_n}} P[a_1,\ldots,a_n]$
are stable if and only if $P$ is stable.
\item $P\nondet Q$, $P\para Q$ are stable if and only if $P$ and $Q$ are stable.
\item $\debutbloc \cond{c_1}{P_1} \ldots \cond{c_n}{P_n} \finbloc$ is stable if and only if for all $i$, $P_i$ is stable. 
\end{enumerate}

\noindent $P$ stable is denoted $P\stable$.
\end{definition}

In the examples, another operator on processes is used: "$\seq$", for sequential composition.
$P\seq Q$ behaves like $Q$ once $P$ has terminated. This require the introduction of a predefined process $\term$, for signaling successful termination. The operator "$;$" can be simulated with "$\para$":
$P\seq Q$ behaves as $(P\para \delta? \prefix Q)\restrictg{\delta}$ where $\delta$ is a fresh gate name and with $\term \procdef \delta! \prefix \stopproc$.

\subsection {Contexts and process states}
\label{subseccontext}

In the inference rules which describe the semantics of processes, the states of processes are process terms $P$ together with contexts $C$, of the form $P\spcontexte C$.
The main purpose of a context is to maintain the quan\-tum state,
stored as $q = \rho$ where $q$ is a sequence of quan\-tum variable names
and $\rho$ a density matrix representing their current quan\-tum state. 
In order to treat classical variables in a similar way, 
modifications of classical variables are allowed. So, for the same reason as in the case of quan\-tum variables, classical values are stored in the context.
Storing and retrieving classical values is represented by functions
$f: \mbox{\it names} \rightarrow\mbox{\it values}$.
The context keeps track of the embedding of variable scopes.
To keep track of parallel composition,
this is done via a "cactus stack" structure of sets of variables, called the environment stack ($s$), which stores variable scopes and types. The set of all the variables in $s$ is denoted $\varpile s$,
"$\pileajout$" adds an element on top of a stack, and "$\pileconcat$" concatenates two stacks.

\begin{definition}
A \textbf{context} is a tuple $\contexteqcq$, where:
\begin{itemize}
\item $s$ is the environment stack;
\item $q$ is a sequence of quan\-tum variable names;
\item $\rho$ is a density matrix representing the quan\-tum state of the variables in $q$;
\item $f$ is the function which associates values with classical variables.
\end{itemize}
\end{definition}

The evaluation of an admissible transformation (rule \ref{inftransfadm} of the semantics) produce a probabilistic result.
This requires the introduction of a probabilistic composition operator for contexts.
This operator is denoted $\symbcontexteprob_p$:
the state $P\spcontexteprobbin{p}{C_1}{C_2}$ is $P\spcontexte{C_1}$ with probability $p$ and $P\spcontexte{C_2}$ with probability $1-p$.
In general, a context is either of the form $\contexteqcq$, or of the form
$\contexteprobqcq$ where the $p_i$'s are probabilities adding to $1$.

As in the case of probabilities introduced by the operator $\probbin p$,
so as to guarantee that probabilistic choice is always solved first, the notion of probabilistic stability for contexts is introduced: a context $C$ is probabilistically stable, which is denoted $C\contextestable$, if it is of the form $\contexteqcq$.
If the context of a process state is not stable, a probabilistic transition must be performed first (rule \ref{infprobcont} of the semantics).

\subsection {Example: teleportation}
\label{subsecteleport}

The teleportation protocol \cite{BennettBrassard93Teleportation} transfers the state of a qubit in a place $A$ into a qubit in a place $B$ with only two classical bits sent from place $A$ to place $B$:
\begin{center}
\begin{minipage}{0.8\textwidth}
\textit {Once upon a time, there were two friends, Alice and Bob who had to separate and live away from each other. Before leaving, each one took a qubit of the same EPR pair.
Then Bob went very far away, to a place that Alice did not know.
Later on, someone gave Alice a mysterious qubit in a state $\ket\psi = \alpha\ket 0+\beta\ket 1$,
with a mission to forward this state to Bob.
Alice could neither meet Bob and give him the qubit, nor clone it and broadcast copies everywhere, nor obtain information about $\alpha$ and $\beta$. Nevertheless, Alice succeeded thanks to the EPR pair and the teleportation protocol.}
\end{minipage}
\end{center}

\noindent
This protocol is described with QPAlg in program \ref{progTeleport}.

\begin{programme}[htbp]
$\begin{array}{rcl}
\eprinit & \procdef & \declparam{\qvar x, \qvar y} \\
&& \tab (( g_1\recep{x} \prefix g_2 \recep y \prefix H[x] \prefix CNot[x,y] \prefix\term)\\
&& \tab \para (g_1\envoi 0\prefix g_2\envoi 0\prefix\term)) \restrictg{g_1,g_2}\\
&&\\
\alice&\procdef& \declparam{\qvar x, \qvar y} \\
&& \tab CNot[x,y] \prefix H[x] \prefix g \envoi{\mstdp{2}{x,y}} \prefix \term\\
&&\\
\bob&\procdef&\declparam{\qvar z}\\
&& \tab \debutbloc \cvar k \findeclvar\\
&& \tab\tab g\recep k\prefix\\
&& \tab\tab \debutbloc \cond{k=0}{I[z]\prefix\term}, \\
&& \tab\tab \cond{k=1}{\sigma_X[z]\prefix\term}, \\
&& \tab\tab \cond{k=2}{\sigma_Z[z]\prefix\term}, \\
&& \tab\tab \cond{k=3}{\sigma_Y[z]\prefix\term} \finbloc \\
&&\tab \finblocse\\
&&\\
\teleport&\procdef& \declparam{\qvar\psi}\\
&& \tab \debutbloc \qvar a, \qvar b \findeclvar\\
&& \tab\tab \eprinit[a,b]\seq\\
&& \tab\tab (\alice[\psi,a]\para\bob[b])\restrictg{g}\\
&& \tab \finblocse\\
\end{array}$
\caption {Implementation of the teleportation protocol}
\label{progTeleport}
\end{programme}

The inference rules can be used to show that this protocol results in Bob's $z$ qubit having the state initially possessed by the $x$ qubit of Alice, with only two classical bits sent from Alice to Bob.

\section {Probabilistic branching bisimilarity}
\label{sectionBisimul}

The operational semantics associates a process graph with a process state. A process graph is a
set of transitions between process states, and an initial state \cite{Fokkink00PA}. The transitions are action transitions: $S\petiteatransition T$ where $S$, $T$ are states, and $a$ is an action (possibly the internal action $\tau$), or probabilistic transitions: $S\ptransition{p} T$, where $p$ is a probability.

In this section, an equivalence relation on process states is defined: probabilistic branching bisimilarity, which identifies states when they are associated with process graphs having the same branching structure.
The bisimilarity defined here is probabilistic because of the probabilistic transitions introduced by quantum measurement and by the operator of probabilistic choice.
The choice of a branching bisimilarity comes from the fact that it abstracts from silent transitions (contrary to strong bisimilarity), but is finer than any other equivalence taking into account silent steps \cite{Glabbeek93eq2}.

This definition is inspired from the definitions in \cite{Fokkink00PA} and \cite{Andova99Art}.

\subsection {Preliminary definitions and notations}

\subsubsection* {Process states}

The set of all possible process states is denoted $\ensetats$.
Let $S, T \in \ensetats$, then $S$ can be written $P\spcontexte{C_P}$ and  $T$, $Q\spcontexte{C_Q}$ where $P$, $Q$ are process terms and $C_P$, $C_Q$ contexts (possibly probabilistic).

Assuming that $S=P\spcontexte{C_P}$ and $C_P=\contexteqcq$,  if $x$ is an initialized qubit in $S$, i.e. $(x,\qubittype)\in s$ and $x\in q$, then $\etatqvar x S$ is the state of $x$ and this state can be obtained with a trace out operation on $\rho$:
$$\etatqvar x S = \traceout{\ensemble x}{q}{\rho}$$ 
 
\subsubsection* {Silent transitions}

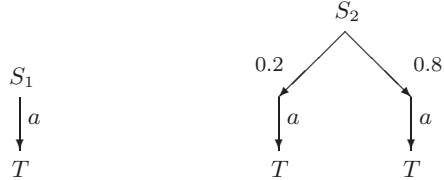
\begin{figure}[htbp]
\begin{center}
\begin{picture}(100,50)
\put(50,35){\vector(0,-1){20}}
\put(53,25){\mbox{$a$}}
\put(46,40){\mbox{$S_1$}}
\put(47,5){\mbox{$T$}}
\end{picture}
\tab
\begin{picture}(100,75)
\put(46,65){\mbox{$S_2$}}
\put(50,60){\vector(1,-1){25}}
\put(50,60){\vector(-1,-1){25}}
\put(16,45){\mbox{\footnotesize 0.2}}
\put(76,45){\mbox{\footnotesize0.8}}
\put(25,35){\vector(0,-1){20}}
\put(75,35){\vector(0,-1){20}}
\put(28,25){\mbox{$a$}}
\put(78,25){\mbox{$a$}}
\put(22,5){\mbox{$T$}}
\put(72,5){\mbox{$T$}}
\end{picture}
\end{center}
\caption {Equivalent states}
\label{figEqStates}
\end{figure}

The transitions considered as silent are of course internal transitions ($\tautransition$) but also probabilistic transitions. The reason is that we want, for example, the states $S_1$ and $S_2$ described in figure \ref{figEqStates} to be equivalent.

Silent transitions will be denoted $\siltransition$. $\seqsiltransition$ stands for a sequence (possibly empty) of silent transitions, and $\siltransition^{0..1}$ stands for zero or one silent transition.

\subsubsection* {Function $\mu$}

Let $\eqrel$ be an equivalence on process states,
$S$ be a process state and $\classeeq S$, its equivalence class with respect to $\eqrel$.
If $M$ is a set of process states, then $S \atteint M$ means that there exists a
sequence of transitions remaining in $M \cup \classeeq{S}$, from $S$ to a state in $M$.

A function $\mu_{\eqrel} : \ensetats \times \parties{\ensetats} \rightarrow [0,1]$ is defined:
$\mu_{\eqrel} : (S,M) \mapsto p$, where $p$ is the probability to reach a state in the set $M$ from a state $S$ without leaving $\classeeq{S} \cup M$.

\begin{figure}
\begin{center}
\begin{tabular}{c}
\begin{picture}(100,50)
\put(50,35){\vector(1,-1){22}}
\put(50,35){\vector(-1,-1){22}}
\put(30,25){\mbox{$a$}}
\put(65,25){\mbox{$b$}}
\put(46,40){\mbox{$S_1$}}
\put(22,5){\mbox{$T$}}
\put(72,5){\mbox{$T$}}
\end{picture}
\end{tabular}
\begin{tabular}{c}
\begin{picture}(100,75)
\put(46,65){\mbox{$S_2$}}
\put(50,60){\vector(1,-1){25}}
\put(50,60){\vector(-1,-1){25}}
\put(16,45){\mbox{\footnotesize 0.5}}
\put(76,45){\mbox{\footnotesize 0.5}}
\put(25,35){\vector(0,-1){20}}
\put(75,35){\vector(0,-1){20}}
\put(28,25){\mbox{$a$}}
\put(78,25){\mbox{$b$}}
\put(22,5){\mbox{$T$}}
\put(72,5){\mbox{$T$}}
\end{picture}
\end{tabular}
\end{center}
\caption {Non equivalent states}
\label{figNonEqStates}
\end{figure}
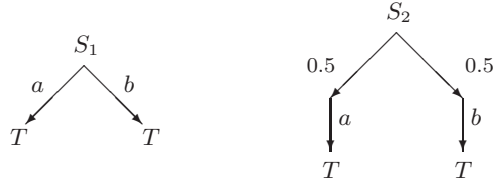

It should be noted that, for this function to yield a probability, nondeterminism must be eliminated in a way which allows the computation of $\mu$.
Here, nondeterminism is treated as equiprobability, but this is just a convention for the definition of  $\mu_{\eqrel}$.
For example, this does not imply the equivalence of the two process states $S_1$ and $S_2$ described in figure \ref{figNonEqStates}.

The function $\mu_{\eqrel}$ is defined as follows:
\begin{itemize}
\item if $S \in M$ then $\mu_{\eqrel}(S,M) = 1$ 

\sautdeligne
\item
\parpic(3cm,6cm)(0cm,2cm)[r][r]{\includegraphics[scale=0.7]{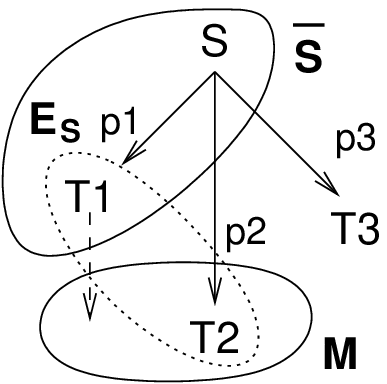}}
else if $\exists\ T \in M \cup \classeeq S$ such that
$S \ptransition p T \atteint M$ then
let $E_S = \ensemble{R\in M\cup\classeeq{S}\ |\ S \ptransition{p_{_R}} R \atteint M}$ in
$$\mu_{\eqrel}(S,M) = \sum_{R\in E_S} p_{_R}\  \mu_{\eqrel}(R,M)$$

\sautdeligne
\item
\parpic(3cm,6cm)(0cm,2.5cm)[r][r]{\includegraphics[scale=0.7]{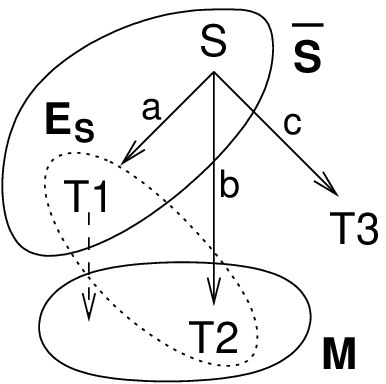}}
else if $\exists\ T \in M \cup \classeeq S$ such that
$S \atransition{a} T \atteint M$ then
let $E_S = \ensemble{R\in M\cup\classeeq{S}\ |\ S \atransition{a_{_{R}}} R \atteint M}$ in
$$\mu_{\eqrel}(S,M) = \frac{1}{\card{E_S}} \sum_{R\in E_S} \mu_{\eqrel}(R,M)$$

\item else $\mu_{\eqrel}(S,M) = 0$ 
\end{itemize}

\subsection {Probabilistic branching bisimulation}

To define bisimilarity, the first step is the definition of a relation of bisimulation on process states. Contrary to the usual definitions of bisimulation, here, a bisimulation has to be an equivalence relation, because of the last point of the definition which ensures that the probability to reach an equivalence class is constant on each equivalence class.

\subsubsection* {Definition}

An equivalence relation is a probabilistic branching bisimulation if and only if:
\begin{itemize}
\parpic(3cm,0cm)(0cm,3.4cm)[r][r]{\includegraphics[scale=0.7]{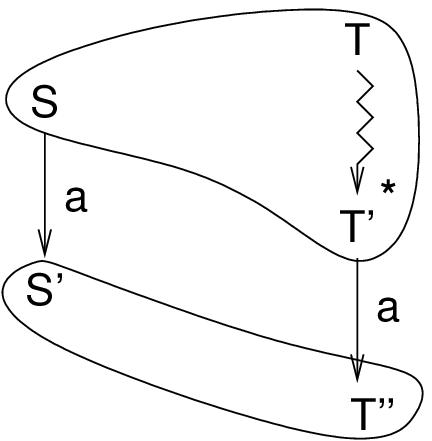}}
\item if $S$ and $T$ are equivalent and if an action $a$ can be performed from $S$, then the same action can be performed from $T$, possibly after several silent transitions;
\item the reached states ($S'$ and $T''$) are equivalent;
\item the action $a$ must occur with the same probability in both cases. The probability to perform an action is the sum for all the branches leading to the action, of the product of the probabilities of each branch. It is calculated thanks to the function $\mu$, defined in the previous section.
\end{itemize}

In the following, $x$, $y$ denote variables and $v$ denotes a classical value.

\sautdeligne
Let $\bisimul$ be an equivalence relation on process states.
$\bisimul$ is a probabilistic branching bisimulation if and only if it satisfies:

\begin{itemize}
\item {\bf Value sending}

if $S \bisimul T$ and $S \atransition{g\envoi v} S'$ then
$\exists\  T', T''$ such that\\
$(T\seqsiltransition T' \atransition{g\envoi v} T'' \ \wedge\  S\bisimul T' \ \wedge\  S'\bisimul T'')$

\item {\bf Qubit sending}

if $S \bisimul T$ and $S \atransition{g\envoi x} S'$ then
$\exists\  T', T''$ such that\\
$(T\seqsiltransition T' \atransition{g\envoi y} T''
\ \wedge\  \etatqvar{x}{S}=\etatqvar{y}{T'}
\ \wedge\  S\bisimul T' \ \wedge\  S'\bisimul T'')$

\item {\bf Value reception}

if $S \bisimul T$ and $S \atransition{g\recep v} S'$ then
$\exists\  T', T''$ such that\\
$(T\seqsiltransition T' \atransition{g\recep v} T''
\ \wedge\  S\bisimul T' \ \wedge\  S'\bisimul T'')$

\item {\bf Qubit reception}

if $S \bisimul T$ and $S \atransition{g\recep x} S'$ then
$\exists\  T', T''$ such that\\
$(T\seqsiltransition T' \atransition{g\recep y} T''
\ \wedge\  \etatqvar{x}{S'}=\etatqvar{y}{T''}
\ \wedge\  S\bisimul T' \ \wedge\  S'\bisimul T'')$

\item {\bf Silent transition}

if $S \bisimul T$ and $S \siltransition S'$ then
$\exists\  T', T''$ such that\\
$(T\seqsiltransition T' \siltransition^{0..1} T''
\ \wedge\  S\bisimul T' \ \wedge\  S'\bisimul T'')$

\item {\bf Probabilities}

if $S \bisimul T$ then $\mu_{\bisimul}(S,M) = \mu_{\bisimul}(T,M),\ \forall M \in \ensetats/\bisimul$
\end{itemize}

\subsubsection* {Bisimulation and recursion}

Because of recursion in a process definition, the computation of $\mu_{\bisimul}$ in the associated process graph can lead to a linear system of equations.
As a consequence, it must be proved that in this case, $\mu_{\bisimul}$ is well-defined and that the system has a unique solution.

Let $S$ be a process state and $M$ be a set of process states, the computation of
$\mu_{\bisimul}(T,M)$ for all $T$ in $\classeeq{S}$ leads to a linear system of equations where the 
$\mu_{\bisimul}(T,M)$ are the unknowns:
$$X = AX + B$$

The $i^{th}$ row in this system can be written:
$x_i = a_{i1} x_1 + \cdots + a_{in} x_n + b_i$.

The coefficients $a_{ij}$ and $b_i$ are either probabilities or average coefficients in the case of nondeterminism. As a consequence: $0 \leq a_{ij} \leq 1$ and $0 \leq b_{i} \leq 1$.
Moreover, in the definition of $\mu_{\bisimul}$, every state in the set $E_S$ is such that there exists a path from that state to the set $M$. 
Therefore, the system of equations obtained can be transformed into a system such that $b_i > 0$, $\forall i \in \llbracket 1,n \rrbracket$.
From now on, we consider that the system verifies this property.

Another property of the system is: $0 \leq \sum_j a_{ij} + b_i \leq 1$, thus $0 \leq \sum_j a_{ij} < 1$.

To prove that the system has a unique solution, it is sufficient to prove $\norme{A}<1$ and use the fixpoint theorem.
The norms for matrices and vectors are:
$$\norme{A} = \sup_{\norme{X}=1} \norme{AX} \hspace{20pt} \norme{X} = \max_{i} \abs{x_i}$$
We obtain:
$$\norme{AX} = \max_{i}\abs{\sum_{j} a_{ij} x_i} \leq \max_i \sum_j (\abs{a_{ij}}\abs{x_i})$$
$$\mbox{and then } \norme{A} \leq \max_i \sum_j \abs{a_{ij}} < 1$$

$\norme{A}<1$ implies that the function $f: X \mapsto AX+B$ is strictly contracting, so
from the fixpoint theorem, we infer that the equation $X =AX+B$ has a unique solution.
Moreover, as $f([0,1]) \subseteq [0,1]$, this solution belongs to $[0,1]$.

As a consequence, the function $\mu_{\bisimul}$ is well-defined even in the case of recursive processes.

\subsection {Probabilistic branching bisimilarity}

\begin{definition}
Two process states $S$ and $T$ are \textbf{bisimilar}, denoted $S \bisimilar T$ if and only if there exists a probabilistic branching bisimulation $\bisimul$ such that $S \bisimul T$.
\end{definition}

\begin{proposition}
Probabilistic branching bisimilarity is an equivalence relation.
\end{proposition}

\begin{preuve}
\begin{description}
\item \textbf{Symmetry.} $S \bisimilar T$ implies that there exists a bisimulation $\bisimul$ such that $S\bisimul T$. Since, $\bisimul$ is an equivalence relation, $T \bisimul S$ and then
$T \bisimilar S$.

\item \textbf{Reflexivity.} The identity relation $\mathcal{I}$ is an equivalence relation which verifies all the points of the definition of the probabilistic branching bisimulation.
$\forall S \in \ensetats$, $S\ \mathcal{I}\ S$, so $S \bisimilar S$.

\item \textbf{Transitivity.}
$S \bisimilar T$ and $T \bisimilar V$, does $S \bisimilar V$? To prove it, we have to find a bisimulation $\bisimul$ such that $S \bisimul T$.

$S \bisimilar T$ and $T \bisimilar V$, so there exist two bisimulations $\bisimul_1$ and $\bisimul_2$ such that $S \bisimul_1 V$ and $V \bisimul_2 T$ (denoted $S \bisimul_1 V\bisimul_2 T$).

Let $\mathtt{Eq}$ give the equivalence closure of a relation and $\circ$ compose two relations:
$S(\bisimul_1\circ\bisimul_2) T$ if and only if there exists $V$ such that
$S\bisimul_1 V\bisimul_2 T$. 
$\bisimul = \mathtt{Eq}(\bisimul_1\circ\bisimul_2)$ is an equivalence relation such that $S \bisimul T$, we prove that it is a bisimulation.

$$S\bisimul T \Leftrightarrow
\begin{array}{|l}
\mbox{either } S = T \mbox{ (reflexive closure)}\\
\mbox{either } \exists V\ |\ S\bisimul_1 V\bisimul_2 T\\
\mbox{either } \exists V\ |\ S\bisimul_2 V\bisimul_1 T \mbox{ (symmetric closure)}\\
\mbox{either } \exists V\ |\ S\bisimul V\bisimul T \mbox{ with } S(\bisimul_1\circ\bisimul_2)V
\mbox{ or } S(\bisimul_2\circ\bisimul_1)V\\
\hspace{25pt}\mbox{(transitive closure)}
\end{array}$$

We develop only the points concerning value sending and probabilities of the definition of a bisimulation, the other points are similar to those developed.

	\begin{description}
	\item \textbf{Value sending.}
	$S \bisimul T$ and $S \petiteatransition S'$ ($a=g\envoi v$), let's prove:
	$\exists\  T', T''$ such that $(T\seqsiltransition T' \petiteatransition T''
	\ \wedge\  S\bisimul T' \ \wedge\  S'\bisimul T'')$.
	
		\begin{description}
		\item {Case $S=T$:} $T'=S$ and $T'' = S'$.
		
		\item {Case $\exists V\ |\ S \bisimul_1 V \bisimul_2 T$:}
		Since $S \bisimul_1 V$, there exist $V', V''$ such that
		$V\seqsiltransition V' \petiteatransition V''$ and
		$S\bisimul_1 V'$ and $S'\bisimul_1 V''$.
		
		\petitsautdeligne
		If $V=V'$, then $V \bisimul_2 T$ implies there exist $T', T''$ such that
		($T\seqsiltransition T' \petiteatransition T''
		\ \wedge\  V\bisimul_2 T' \ \wedge\  V''\bisimul_2 T'')$.\\
		$S\bisimul_1 V$ and $V \bisimul_2 T'$, so $S\bisimul T'$.
		$S'\bisimul_1 V''$ and $V'' \bisimul_2 T''$, so $S' \bisimul T''$.
		
		\petitsautdeligne
		Otherwise $V\siltransition V_1\siltransition\cdots\siltransition V_n\siltransition 
		V''\petiteatransition V'$.
		By applying the point on silent transition of the definition of a bisimulation to
		the relation $\bisimul_2$ with the successive $V_i$, we obtain:
		$\exists\  T', T''$ such that $(T\seqsiltransition T' \petiteatransition T''
		\ \wedge\  S\bisimul T' \ \wedge\  S'\bisimul T'')$.
			
		\item {Case $\exists V\ |\ S \bisimul_2 V \bisimul_1 T$:} idem previous case.
		
		\item {Case $\exists V\ |\ S \bisimul V \bisimul T$ with
		$S(\bisimul_1\circ\bisimul_2)V$ or $S(\bisimul_2\circ\bisimul_1)V$:} by induction on the
		sequence of relations.
		\end{description}
	
	\item \textbf{Probabilities.}
	Let $\ensemble{C_i}_{i\in I}$, $\ensemble{D_j}_{j\in J}$ and $\ensemble{M_k}_{k\in K}$ be
	the sets of equivalence classes of respectively $\bisimul_1$, $\bisimul_2$ and $\bisimul$.
	 
	For all $k$, $l$ in $K$, we prove that for all $S, T \in M_k$,
	$\mu_\bisimul(S,M_l)=\mu_\bisimul(T,M_l)$.
	
	Firstly, we need to know the relations between the $M_k$'s and the $C_i$'s and $D_j$'s.
	If $S,T \in C_i$, then $S \in M_k$ implies $T \in M_k$, because $S \bisimul_1 T \bisimul_2 T$.
	As a consequence $\forall i \in I$, $\forall k \in K$, $C_i \subseteq M_k$ or
	$C_i \cap M_k = \emptyset$.
	Since the equivalence classes of an equivalence relation form a partition of the space
	considered, $M_k = \bigcup_{i\in I_k} C_i$.
	Similarly for $\bisimul_2$: $M_k = \bigcup_{j\in J_k} D_j$.
	Moreover, for all $S$, $T$ in $M_k$, there exists a path $V_1, \ldots, V_p$ such that
	$S \bisimul_1 V_1 \bisimul_2 V_2 \bisimul_1 \ldots \bisimul_2 V_p \bisimul_1 T$.
	
	$$M_k = \bigcup_{i\in I_k} C_i = \bigcup_{j\in J_k} D_j \tab
	M_l = \bigcup_{i\in I_l} C_i = \bigcup_{j\in J_l} D_j$$
	
	Now, let's compute $\mu_\bisimul$ as a function of $\mu_{\bisimul_1}$
	(cf. the example at the end of this section).
	Let $S \in M_k$, there exists $n$ such that $S \in C_n$.
	$\mu_{\bisimul}(S,M_l)$  is the probability to reach
	$M_l$ from $S$ without leaving $M_k \cup M_l$, in other words, it is the sum of the
	probabilities to reach each $C_j$ for $j \in I_l$ without leaving
	$\bigcup_{i \in I_k\cup I_l} C_i$.
	
	Since $\bisimul_1$ is a bisimulation, the probability to reach $C_j$ from $C_i$ is constant for
	all state in $C_i$, this probability will be denoted $\mu_{\bisimul_1}(C_i, C_j)$.
	
	$$\mu_{\bisimul}(S,M_l) = \sum_{j \in I_l} \mu_{\bisimul} (S,C_j) =
	\sum_{j \in I_l} \tilde\mu_{\bisimul} (C_n,C_j)$$ 
	with $\tilde{\mu}_\bisimul$ defined by:
	$$\begin{array}{rcl} 
	\tilde{\mu}_\bisimul(C_i,C_i) & = & 1\\
	\tilde{\mu}_\bisimul(C_i,C_j) & = & \sum_{m\in E_j} \mu_{\bisimul_1}(C_i,C_m) 
	\tilde{\mu}_\bisimul(C_m,C_j)
	\end{array}$$
	where $E_j = \ensemble{m\ |\ m\in I_k \mbox{, and }C_m \atteint C_j}$.
	
	This shows that the function $S \mapsto \mu_{\bisimul}(S, M_l)$ is constant on $C_i$, for all
	$i \in I_k$. Similarly, this function is also constant on $D_j$, for all $j$ in $J_k$.
	
	With these properties:
		\begin{itemize}
		\item if $S, T \in C_n$, then $\mu_\bisimul(S,M_l) = \mu_\bisimul(T,M_l)$
		
		\item if $S, T \in D_m$, $\mu_\bisimul(S,M_l) = \mu_\bisimul(T,M_l)$

		\item else there exists $\ensemble{V_1,\ldots,V_p}$ such that
		$\forall i \in \llbracket 1,p\rrbracket, V_i \subseteq M_k$,\\
		and $S \bisimul_1 V_1 \bisimul_2 V_2 \bisimul_1 \ldots \bisimul_2 V_p \bisimul_1 T$.
		$$\begin{array}{l}
		\mu_\bisimul(S,M_l) = \mu_\bisimul(V_1,M_l),\\
		\mu_\bisimul(V_1,M_l) = \mu_\bisimul(V_2,M_l),\\
		\cdots\\
		\mu_\bisimul(V_p,M_l) = \mu_\bisimul(T,M_l)
		\end{array}$$
		As a consequence, $\mu_\bisimul(S,M_l) = \mu_\bisimul(T,M_l)$.
		\end{itemize}
	\end{description}
\end{description}
\end{preuve}

\paragraph {Example of computation of $\mu_{\eqrel}$ as a function of $\mu_{\eqrel_1}$.}

\begin{figure}[h]
\centering
\includegraphics[scale=0.6]{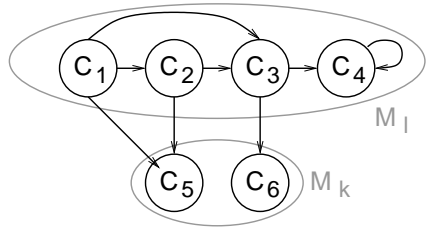}
\caption{Example of equivalence classes}
\label{figclasseseq}
\end{figure}

Figure \ref{figclasseseq} presents the equivalence classes on a process graph of two relations: $\eqrel$ which equivalence classes are $M_l$ and $M_k$ and $\eqrel_1$ with the $C_i$'s as equivalence classes. The arrows represent transitions of the process graph.

Let $S \in C_1$.
$$\begin{array}{rcl}
\mu_{\eqrel}(S,M_l) & = & \mu_{\eqrel}(S,C_4) + \mu_{\eqrel}(S,C_5)\\
&&\\
\mu_{\eqrel}(S,C_4) & = & \tilde\mu_{\eqrel}(C_1,C_4)\\
& = & \mu_{\eqrel_1}(C_1,C_4) + \mu_{\eqrel_1}(C_1,C_2) \tilde\mu_{\eqrel}(C_2,C_4)\\
& = & \mu_{\eqrel_1}(C_1,C_4) + \mu_{\eqrel_1}(C_1,C_2) \mu_{\eqrel_1}(C_2,C_4)\\
&&\\
\mu_{\eqrel}(S,C_5) & = & \mu_{\eqrel_1}(C_1,C_2) \mu_{\eqrel_1}(C_2,C_5)\\
\end{array}$$

We obtain:
$$\mu_{\eqrel}(S,M_l) =  \mu_{\eqrel_1}(C_1,C_4) + \mu_{\eqrel_1}(C_1,C_2)(\mu_{\eqrel_1}(C_2,C_4) + \mu_{\eqrel_1}(C_2,C_5))$$

\section {Example: "H $\bisimilar$ its measurement-based simulation?"}
\label{sectionHad}

Quantum measurement is universal for quantum computation \cite{Nielsen01QCMeas}.
This means that every unitary transformation can be simulated using mea\-su\-re\-ments only.
We are interesting in proving with QPAlg that a unitary transformation and its mea\-su\-re\-ment-based simulation behave the same way. We consider the case of the Ha\-da\-mard transformation $H$.

\subsection {Measurement-based simulation of $H$}

There exist several models of quantum computation via measurements only.
We use the model based on state transfer defined by Perdrix in \cite{Perdrix04StateTransfer}.

\begin{figure}[hbtp]
\begin{center}
\includegraphics[scale=0.55]{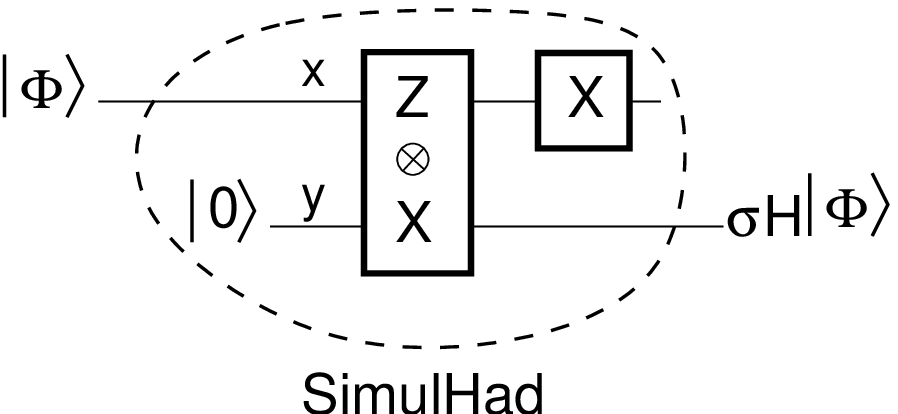}
\includegraphics[scale=0.55]{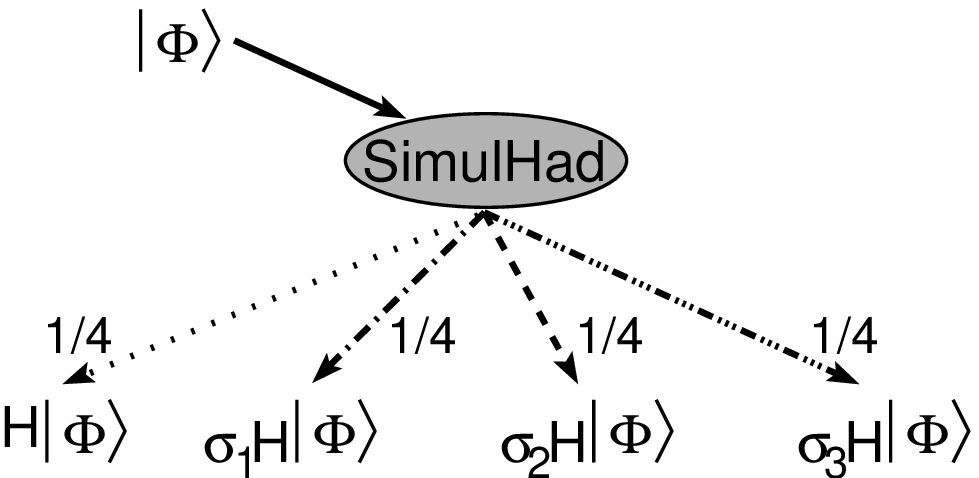}
\end{center}
\caption{A step of simulation of $H$}
\label{figSimulH}
\end{figure}

\begin{figure}[hbtp]
\begin{center}
\includegraphics[scale=0.6]{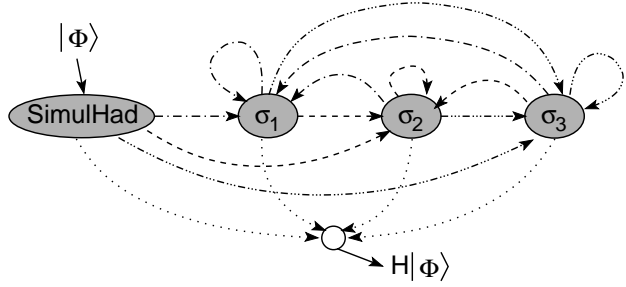}
\end{center}
\caption{Full simulation of $H$}
\label{figFullSimulH}
\end{figure}

The gate network in figure \ref{figSimulH} describes a step of simulation of $H$.
This simulation needs an auxiliary qubit initialized in the state $\ket 0$, it consists in two measurements: one on two qubits with observable $Z\otimes X$ and the other on one qubit with observable $X$. $X$ and $Z$ are Pauli observables.
This step simulates $H$ up to a Pauli operator $\sigma$. If $\sigma = I$, $H$ has been simulated, otherwise, a correction must be applied to the result. This correction consists in simulating $\sigma$ (Pauli operators are their proper inverses) in the same way as $H$ has been simulated. The full simulation of $H$ is given by the automaton on figure \ref{figFullSimulH}. The states \textbf{SimulHad}, $\pmb{\sigma_1}$, $\pmb{\sigma_2}$ and $\pmb{\sigma_3}$ represent a step of simulation of $H$ and of the Pauli operators.

\subsection {Modelling with QPAlg}

\paragraph {Unitary transformation of Hadamard.}
The process $\proc{Had}$ applies the Hadamard transformation on a qubit received from gate $g$ and returns this qubit through gate $h$.
$$\proc{Had} \procdef \debutbloc \qvar{x} \findeclvar
g \recep x \prefix H[x] \prefix h \envoi x \prefix \stopproc \finbloc$$

\paragraph {Simulation of Hadamard}
As for the process $\proc{Had}$, the process $\proc{SimulHad}$ (program \ref{progSimulHad}) begins by receiving a qubit $x$ from gate $g$. Then the Hadamard transformation is simulated in the way described by the automaton shown in figure \ref{figFullSimulH}.
At the end, the result is in $y$ (and not $x$) which is sent through gate $h$.
$\pmb{\sigma}$ is a process simulating a Pauli transformation, the Pauli transformation simulated is specified by the parameter $i$ in $\sigma[i,y]$.
The process $\pmb{\sigma}$ is not given, it is organized in a way similar to $\proc{SimulHad}$.


\begin{programme}[hbtp]
$\proc{SimulHad}  \procdef$

$\begin{array}{ll}
\debutbloc \qvar{y},\ \cvar{i} \findeclvar & \\
\tab \debutbloc \qvar{x},\ \cvar{m, n} \findeclvar & \\
\tab\tab \gate{g} \recep x \prefix
& \comment{initialization of $x$}\\
\tab\tab (\gate{p} \recep y \prefix \term \para \gate{p} \envoi 0 \prefix \term)
\restrictg{\gate{p}}\seq
& \comment{initialization of $y$ to $\ket0$}\\
\multicolumn{2}{l}
{\tab\tab (q_1 \envoi{Z\otimes X}[x,y]\prefix q_2 \envoi{X[x]}\prefix\term
\para q_1 \recep m \prefix q_2 \recep n \prefix \term)
\restrictg{q_1,q_2}\seq}\\
\tab\tab (\debutbloc\\
\tab\tab\tab \cond{(m=1 \wedge n=1)}{\gate{r} \envoi 0\prefix \term},\\
\tab\tab\tab \cond{(m=1 \wedge n=-1)}{\gate{r} \envoi 1\prefix \term},
& \comment{initialization of $i$}\\
\tab\tab\tab \cond{(m=-1 \wedge n=1)}{\gate{r} \envoi 3\prefix \term},
& \comment{depending on}\\
\tab\tab\tab \cond{(m=-1 \wedge n=-1)}{\gate{r} \envoi 2\prefix \term}
& \comment{measurement results}\\
\tab\tab \finblocse \para \gate{r} \recep{i}\prefix \term) \restrictg{\gate{r}}\\
\tab \finblocse \seq \pmb{\sigma}[i,y] \seq 
& \comment{correction} \\
\tab h \envoi{y} \prefix \stopproc
& \comment{result sending} \\
\finblocse
\end{array}$
\caption{Process for the simulation of $H$}
\label{progSimulHad}
\end{programme}

\begin{figure}
\begin{center}
\includegraphics[scale=0.6]{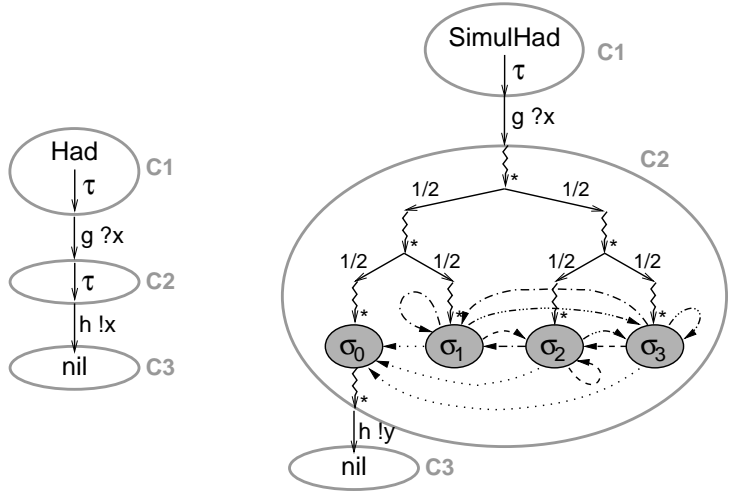}
\end{center}
\caption{Process graphs of $\proc{Had}$ and $\proc{SimulHad}$ in an empty context}
\label{figPGHad}
\end{figure}

The operational semantics associates with the processes $\proc{Had}$ and $\proc{SimulHad}$ (program \ref{progSimulHad}) in empty contexts the process graphs described in figure \ref{figPGHad}.

\subsection {Bisimilarity}

To prove that the application of the Hadamard transformation and its measure\-ment-based simulation have the same behavior, we prove that their modellings in QPAlg are bisimilar, {\it i.e.} that there exists a bisimulation between $\proc{Had}$ and $\proc{SimulHad}$ in empty contexts.

The process graphs of $\proc{Had}$ and $\proc{SimulHad}$ in empty contexts are given in figure \ref{figPGHad}. $C_1$, $C_2$, $C_3$ and $C_4$ represents equivalence classes of an equivalence relation $\eqrel$ on process states. This relation is a bisimulation,
the main points of this proof are:
\begin{itemize}
\item from each process state in $C_1$ the transition $g \recep x$ can be performed, possibly after several silent transitions
\item from each process state in $C_2$ the transition $h \envoi x$ or $h \envoi y$ can be performed, possibly after several silent transitions
\item when $h \envoi x$ and $h \envoi y$ are performed, the states of $x$ and $y$ are the same
\item the probability to reach $C_3$ from each state in $C_2$ must be the same, which corresponds for any state $S$ in $C_2$, to $\mu_\eqrel (S,C_3) = 1$
\end{itemize}

\begin{figure}
\begin{tabular}{c}
\includegraphics[scale=0.6]{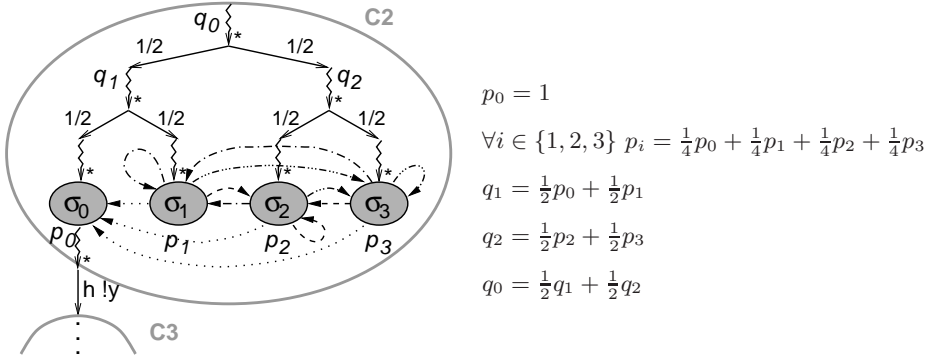}
\end{tabular}$\begin{array}{l}
p_0 = 1 \\
\forall i \in \ensemble{1,2,3}\ 
p_i = \frac{1}{4} p_0 + \frac{1}{4} p_1 + \frac{1}{4} p_2 + \frac{1}{4} p_3\\
q_1 = \frac{1}{2} p_0 + \frac{1}{2} p_1\\
q_2 = \frac{1}{2} p_2 + \frac{1}{2} p_3\\
q_0 = \frac{1}{2} q_1 + \frac{1}{2} q_2\\
\end{array}$
\caption{Computation of the probabilities}
\label{figSystEq}
\end{figure}

As regards the last point, the computation of the function $\mu_\eqrel$ on each state in $C_2$ leads to the system of equations in figure \ref{figSystEq}.
We obtain $\forall i \in \ensemble{0,\ldots,3}$, $p_i = 1$, and then $q_0=q_1=q_2=1$.
From any state in the class $C_2$, the probability to reach a state in the class $C_3$ is 1.

\renewcommand{\arraystretch}{1.5}
\renewcommand{\arraystretch}{1}

\sautdeligne
To conclude, there exists a bisimulation relation between the modelling of the application of Hadamard and the modelling of its measurement-based simulation.

\section {Is $\bisimilar$ a congruence?}
\label{sectionCongru}

A natural question about the bisimilarity defined in this section is: is it a congruence?
In fact, this question does not make much sense since this relation is defined on process states (process and context) and there exist no composition operators on process states.

There are two possibilities to make it a congruence, either the operators on processes are extended to operators on process states or a congruence on processes is defined from the bisimilarity relation on process states.

We explore the second possibility. A relation $\congrusem$ on processes is defined:
if $P$ and $Q$ are two processes,
$P \congrusem Q \Leftrightarrow \forall C,P\spcontexte{C} \bisimilar Q\spcontexte{C}$.

Nonetheless, $\congrusem$ is not a congruence, as shown by the two following examples:
$$\begin{array}{rrcl}
& H[x]\prefix g\envoi x\prefix \stopproc & \congrusem & g\envoi x\prefix \stopproc\\
\mbox{but} & H[x]\prefix g\envoi x \prefix \stopproc \nondet h\envoi 2\prefix \stopproc
& \not\congrusem &
g\envoi x\prefix \stopproc \nondet h\envoi 2\prefix \stopproc\\
&&&\\
& H[x]\prefix g\envoi x\prefix \stopproc
& \congrusem &
g\envoi x\prefix \stopproc \probbin{0.3} g\envoi x\prefix \stopproc\\
\mbox{but} & H[x]\prefix g\envoi x\prefix\stopproc\nondet h\envoi 2\prefix\stopproc
& \not\congrusem&
(g\envoi x\prefix\stopproc\probbin{0.3} g\envoi x\prefix\stopproc)\nondet h\envoi 2\prefix \stopproc
\end{array}$$

To overcome these problems, probabilistic rooted branching bisimulation and probabilistic rooted branching bisimilarity are defined. 

\begin{definition}
Let $\bisimul$ be an equivalence relation.
$\bisimul$ is a \textbf{probabilistic rooted branching bisimulation} if and only if:
\begin{itemize}
\item 
if $S \bisimul T$ and $S \petiteatransition S'$ ($a = g\envoi v $ or $g\recep v$ or $\tau$) then
$T \petiteatransition T'$ with $S'\bisimilar T'$
\item 
if $S \bisimul T$ and $S \atransition{g\envoi x} S'$,then
$T \atransition{g\envoi y} T'$ with $\etatqvar{x}{S}=\etatqvar{y}{T}$ and $S'\bisimilar T'$
\item 
if $S \bisimul T$ and $S \atransition{g\recep x} S'$, then
$T\atransition{g\recep y} T'$ with $\etatqvar{x}{S'}=\etatqvar{y}{T'}$ and $S'\bisimilar T'$
\item 
if $S \bisimul T$ and $S \ptransition{p} S'$, then
$T \ptransition{q} T'$ with $S' \bisimul T'$
\item 
if $S \bisimul T$ then $\mu_{\bisimilar}(S,M) = \mu_{\bisimilar}(T,M),
\ \forall M \in \ensetats/\bisimilar$
\end{itemize}
\end{definition}

\begin{definition}
Two process states $S$ and $T$ are \textbf{probabilistic rooted branching bisimilar}, denoted $S \rootedbisimilar T$, if and only if there is a probabilistic branching bisimulation $\bisimul$ such that $S \bisimul T$.
\end{definition}

\begin{proposition}
$\rootedbisimilar$ is an equivalence relation.
\end{proposition}

\begin{preuve}
From the fact that $\bisimilar$ is an equivalence relation.
\end{preuve}

\sautdeligne
\begin{remarque}
For all $S$, $T$ in $\ensetats$: $S\rootedbisimilar T$ implies $S \bisimilar T$
\end{remarque}

\begin{definition}
Let $P$ and $Q$ be two processes.
$$P \congrusem Q \Leftrightarrow \forall C,P\spcontexte{C} \rootedbisimilar Q\spcontexte{C}$$
\end{definition}

\begin{proposition}
$\congrusem$ is an equivalence relation and is preserved by variable declaration, action prefix, nondeterministic choice, probabilistic choice, conditional choice and restriction.
\end{proposition}

\begin{preuve}
Since $\rootedbisimilar$ is an equivalence relation, it is easy to see that $\congrusem$ is also an equivalence relation.

Let $P$, $Q$ be processes such that $P\congrusem Q$, that is to say $\forall C,P\spcontexte{C} \rootedbisimilar Q\spcontexte{C}$.
\begin{description}
\item \textbf {Action prefix.} Let $a$ be an action.
The question is: are $a\prefix P \spcontexte{C}$ and $a\prefix Q \spcontexte{C}$ probabilistically rooted branching bisimilar for all context $C$?

If $C$ is probabilistically stable:
$a\prefix P \spcontexte{C} \petiteatransition P \spcontexte{C'}$ and
$a\prefix Q \spcontexte{C} \petiteatransition Q \spcontexte{C'}$.
$P\spcontexte{C'} \rootedbisimilar Q\spcontexte{C'}$ implies
$P\spcontexte{C'} \bisimilar Q\spcontexte{C'}$.
$\bisimul = \ensemble{(a\prefix P \spcontexte{C},a\prefix Q \spcontexte{C})}$ is a probabilistic rooted branching bisimulation.

If $C = \pcontexteprob{p_i}{C_i}$: for all $i$,
$a\prefix P \spcontexte{C} \ptransition{p_i} a\prefix P \spcontexte{C_i}$ and
$a\prefix Q \spcontexte{C} \ptransition{p_i} a\prefix Q \spcontexte{C_i}$.
From the previous case, $\forall\ i$:
$a\prefix P \spcontexte{C_i} \rootedbisimilar a\prefix Q \spcontexte{C_i}$,
we deduce $a\prefix P \spcontexte{C} \rootedbisimilar a\prefix Q \spcontexte{C}$.

\item \textbf {Nondeterminism.}
Let $R$ be a process. The question is: are $P\nondet R\spcontexte C$ and $Q\nondet R\spcontexte{C}$ probabilistic rooted branching bisimilar, for all context $C$?

If $C$ is probabilistically stable:
If $P\nondet R\spcontexte{C} \divtransition P'\spcontexte{C_P}$ then, as $P\spcontexte C \rootedbisimilar Q \spcontexte C$, there exists $Q'\spcontexte{C_Q}$ such that $Q \nondet R \spcontexte{C} \divtransition Q'\spcontexte{C_Q}$ and
$P'\spcontexte{C_P} \bisimilar Q'\spcontexte{C_Q}$.
Otherwise $P \nondet R\spcontexte{C} \divtransition R'\spcontexte{C''}$ then,
$Q \nondet R \spcontexte{C} \divtransition R'\spcontexte{C''}$ and
$R'\spcontexte{C''} \bisimilar R'\spcontexte{C''}$.

If $C$ is not probabilistically stable, we reduce the problem to the previous point after a probabilistic transition.
\end{description}

The other points are similar to those developed.
\end{preuve}

\sautdeligne
We conclude that $\congrusem$ is preserved by all operators except $\para$, as shown by the following example:
$(g \envoi 2 \prefix h \envoi 3 \prefix \stopproc \para g \recep x \prefix h \recep x \prefix \stopproc)
\restrictg{g,h} \congrusem (h \envoi 3 \prefix \stopproc \para h \recep x \prefix \stopproc)
\restrictg{h}$, nonetheless, $(g \envoi 2 \prefix h \envoi 3 \prefix \stopproc \para g \recep x \prefix h \recep x \prefix \stopproc) \restrictg{g,h} \para k \envoi x \prefix \stopproc \not \congrusem (h \envoi 3 \prefix \stopproc \para h \recep x \prefix \stopproc) \restrictg{h} \para k \envoi x \prefix \stopproc$.
The left process can send $2$ or $3$ through gate $k$ whereas the right process can only send $3$.

This problem could be overcome by restricting processes in parallel not to use the same variable names. This is done in CQP \cite{NagGay04Turku} and can be justified by the fact that variables cannot be at two places at the same time. However, because of entanglement in the quantum state, this does not solve the whole problem:
\[\begin{array}{rcl}
\mstdp{1}{x} \prefix \stopproc  & \congrusem &
I[x] \prefix (0\prefix \stopproc \probbin{0.5} 1 \prefix \stopproc) \\
\mstdp{1}{x} \prefix \stopproc \para g \envoi y \prefix \stopproc & \not \congrusem &
(I[x] \prefix (0\prefix \stopproc \probbin{0.5} 1 \prefix \stopproc)) \para g \envoi y \prefix \stopproc
\end{array}\]
In a configuration where the state of $x$ and $y$ is $\frac{1}{\sqrt 2}(\ket{00}+\ket{11})$ (EPR state), the left process can send the qubit $y$ in the mixed state $\ensemble{0.5: \ket{0}, 0.5:\ket{1}}$ through gate $g$ if $\mstdp 1 x$ has not been applied, or in the states $\ket 0$ or $\ket 1$ if the measurement has been applied, whereas the right process can only send $y$ in the state $\ensemble{0.5: \ket{0}, 0.5:\ket{1}}$.

\section {Conclusion}

This paper has presented a process algebra for quantum programming which can describe both classical and quantum programming, and their cooperation. This language has an operational semantics, one of its peculiarities is the introduction of probabilistic transitions, due to quantum measurement and to the operator of probabilistic choice.

Then a semantical equivalence relation on process states has been defined. This equivalence is a bisimulation which identifies processes associated with process graphs having the same branching structure. From this bisimulation, an equivalence relation on processes has been defined. This relation is preserved by all the operators of the process algebra except parallel composition. This is a first step toward the verification of quantum cryptographic protocols.

Several extensions are possible. As already mentioned, we could define a congruence on process states from the bisimulation defined here by extending the operators on processes to operators on process states. Another possible extension is the definition of a type system to verify statically properties such as the no-cloning theorem (quantum variables cannot be copied).




\appendix

\section{The quantum process algebra}
\label{annexQPA}

\subsection{Syntax}
\label{annexQPASyntax}

\begin{supertabular}{lcl}
\bf Expressions\\
\it qexp & $\syntaxdef$ & \it qvar $\pmb \otimes$ qexp $|$ qvar\\
&&\\
\it nexp & $\syntaxdef$ & \it nfact $\pmb +$ nexp $|$ nfact $\pmb -$ nexp $|$ nfact\\
\it nfact & $\syntaxdef$ & \it nterm $\pmb \times$ nfact $|$ nterm $\pmb \div$ nfact $|$ nterm\\
\it nterm & $\syntaxdef$ & \it nvar $|$ nval $|$ transf\_admissible $\pmb [$ qexp $\pmb ]$\\
	&$|$& $\pmb ($ nexp $\pmb )$ $|$ $\pmb -$nterm\\
&&\\
\bf Actions\\
\it communication & $\syntaxdef$ & \it gate {\bf !} variable $|$ gate {\bf !} nexp\\
	&$|$& \it  gate {\bf ?} variable\\
\it action & $\syntaxdef$ & \it communication $|$ nexp \\
&&\\
\bf Processes\\
\it type & $\syntaxdef$ &\it $\pmb{\nattype}\ |\ \pmb\qubittype$\\
\it variable\_decl & $\syntaxdef$ &
	\it variable $\pmb :$ type $\{\pmb ,$ variable $\pmb :$ type $\}^*$ \\
&&\\
\it process & $\syntaxdef$ & \it $\pmb\stopproc$ \\
                  &$|$& \it name $\{\pmb[$ variable\_list $\pmb]\}$ \\
                  &$|$& \it param\_decl $\pmb\bullet$ process $\pmb[$ variable\_list $\pmb]$ \\
                  &$|$& \it $\pmb\debutbloc$ variable\_decl $\pmb .$ process $\pmb\finbloc$ \\
                  &$|$& \it action $\pmb\prefix$ process\\
                  &$|$& \it  process $\pmb\para$ process\\
                  &$|$& \it  process $\pmb\nondet$ process\\
                  &$|$& \it  $\pmb\debutbloc$ condition $\pmb\rightarrow$ process
                  $\{\pmb ,$ condition $\pmb\rightarrow$ process $\}^* \pmb\finbloc$ \\
                  &$|$& \it  process $\{\pmb\oplus_{\mbox{probability}}$ process$\}^+$\\
                  &$|$& \it  process $\pmb \backslash \pmb\{$ gate\_list $\pmb\}$ \\
&&\\
\it param\_decl & $\syntaxdef$ &
	\it $\pmb \debutbloc$ variable\_decl $\pmb\finbloc$\\
&&\\
\it process\_decl & $\syntaxdef$ & \it name $\pmb\procdef$ process
	$|$ \it nom $\pmb\procdef$ param\_decl $\pmb\bullet$ process\\
\end{supertabular}

\subsection{Main inference rules of the semantics}
\label{annexQPASem}

The semantics is specified with inference rules which give the evolution of the states of processes. There are five kinds of transitions:
\begin{itemize}
\item transitions for evaluating expressions: $\vtransition$ and $\etransition$
\item action transitions: $\atransition\alpha$ where $\alpha$ is $g\envoi x$ or $g\recep x$;
\item silent transitions: $\tautransition$, for internal transition;
\item probabilistic transitions: $\ptransition p$, where $p$ is a probability.
\end{itemize}

In the following, $P, Q, P', Q', P_i$ and $P_i'$ are processes, $C$, $C'$ and $C_i$ are contexts, $\alpha$ is an action, $g$ is a communication gate, $v$ is a value, $x$ is a variable, and $c_j$ is a condition.

\subsubsection*{Expressions}
\regleinfcond{infvc}{}{x\scontexteqcq \vtransition f(x)\scontexteqcq}{x\in\dom{f}}

\regleinf{inftransfadm}{}
{A[x_1\otimes\ldots\otimes x_n] \spcontexte C \vtransition y \spcontexteprob{p_i}{C_i}}
avec
\begin{itemize}
\item $A = \ensemble{A_{\tau_1},\ldots,A_{\tau_n}}$, admissible transformation
\item $y$, fresh variable name
\item $C = \contexte{e\pileajout s}{q}{\rho}{f}$
\item $C_i = \contexte{(\ensemble{y:\nattype}\cup e)\pileajout s}{q}{\rho_i}
{f\cup\ensemble{y\mapsto\tau_i}}$
\item $x_1,\ldots,x_n \in q $
\item $p_i = \trace{\applip{A_{\tau_i}}{\rho}}$ and $\rho_i = \frac{1}{p_i} \applip{A_{\tau_i}}{\rho}$
\end{itemize}

\subsubsection*{Evaluation contexts}

$E[\ ]$ is an evaluation context of an expression, it is an expression in which a sub-expression has been replaced by $[\ ]$. Similarly, $F[\ ]$ is an evaluation context of a process.

\regleinf{infcontexp}{e \spcontexte{C} \vtransition e' \spcontexte{C'}}
{E[e] \spcontexte{C} \etransition E[e'] \spcontexte{C'}}

\regleinf{infconteval}{e \spcontexte{C} \etransition e' \spcontexte{C'}}
{F[e] \spcontexte{C} \tautransition F[e'] \spcontexte{C'}}

\subsubsection*{Action Prefix}
\paragraph{\em Communication.}
\regleinfcond{infenvoivaleur}
{}{g\envoi{v} \prefix P \spcontexte{C} \atransition{g\envoi{v}} P \spcontexte{C}}{v \in \n}
\regleinfcond{infenvoiqubit}
{}{g\envoi{x} \prefix P \spcontexte{C} \atransition{g\envoi{x}} P \spcontexte{C}} 
{C=\contexteqcq \mbox{ and } x\in q}

For all $v \in \n$:
\regleinf{infrecoitcl}{}{g\recep x \prefix P \spcontexte C \atransition{g\recep v} P \spcontexte C'}
with $C = \contexteqcq$, $C' = \contexte{s}{q}{\rho}{f\cup\ensemble{x\mapsto v}}$
and $(x, \nattype) \in s$

\sautdeligne
For all density matrix $\nu$ of dimension $2$:
\regleinf{infrecoitq}{}{g\recep x \prefix P \spcontexte C \atransition{g\recep x} P \spcontexte C'}
with
\begin{itemize}
\item $C = \contexteqcq$, $C'=\contexte{s}{x.q}{\nu\otimes\rho}{f}$
\item $(x,\qubittype)\in s$ and $x\not\in q$
\end{itemize}

\paragraph{\em Expression.}
\regleinfcond{infprefexp}{}{v \prefix P \spcontexte C \tautransition P \spcontexte C}
{v \in \n}

\subsubsection*{Probabilities}
\regleinfcond{infprobcont}
{}{P \spcontexteprob{p_j}{C_j} \ptransition{p_i} P\spcontexte{C_i}}
{\sum_j p_j = 1}
\regleinfcond{infprobproc}
{}{\prob{p_j} P_j \spcontexte{C} \ptransition{p_i} P_i\spcontexte{C}}
{\sum_j p_j = 1}

\subsubsection *{Nondeterministic choice}
\regleinfcond{infnondetdiv}
{P \spcontexte{C} \divtransition P'\spcontexte{C'}}
{P \nondet Q \spcontexte{C} \divtransition P'\spcontexte{C'}}
{P\stable, Q\stable}
where $\divtransition$ represents any transition.

\regleinfcond{infnondetprob1}
{P \spcontexte{C} \ptransition p P' \spcontexte{C'}}
{P \nondet Q \spcontexte{C} \ptransition{p} P'\nondet Q \spcontexte{C'}}
{Q\stable}

\regleinf{infnondetprob2}
{P \spcontexte{C} \ptransition p P' \spcontexte{C'}\hspace{20pt}
Q \spcontexte{C} \ptransition q Q' \spcontexte{C''}}
{P \nondet Q \spcontexte{C} \ptransition{pq} P'\nondet Q' \spcontexte{C}}

\subsubsection*{Parallel composition}

In the rules for parallel composition, $C$, $C_P$ and $C_Q$ are defined as:
\begin{itemize}
\item $C = \contexte{\pileajoutp s {(s_P\para s_Q)}}{q}{\rho}{f}$
\item $C_P = \contexte{s_P\pileconcat s}{q}{\rho}{f}$
\item $C_Q = \contexte{s_Q\pileconcat s}{q}{\rho}{f}$
\end{itemize}

In the definition of $C$, the operator $\para$ permits to build a cactus stack (see paragraph \ref{subsecQuantVar}).
In the cactus stack $\pileajoutp s {(s_P\para s_Q)}$ of the process $P\para Q$, the names in $s$
correspond to variables shared by $P$ and $Q$ whereas the names in $s_P$ (resp. $s_Q$) correspond to variables declared in $P$ (resp. $Q$).
\regleinfcond{infcomppara}
{P \spcontexte{C_P} \atransition{\alpha} P' \spcontexte{C_P'}}
{P \para Q \spcontexte{C} \atransition{\alpha} P' \para Q \spcontexte{C'}}
{P\stable, Q\stable}
where
\begin{itemize}
\item If $C_P' =\contexte{s'}{q'}{\rho'}{f'}$ then 
$C' = \contexte{\pileajoutp s {(s_P'\para s_Q)}}{q'}{\rho'}{f'}$
with $s_P'$ such that $s' = s_P'\pileconcat s$ ($P$ can neither add to nor remove variables from $s$)

\item If $C_P' =\contexteprob{p_i}{s_i'}{q_i'}{\rho_i'}{f_i'}$
then $C' = \contexteprob{p_i}{\pileajoutp s {({s_P}_i'\para s_Q)}}{q_i'}{\rho_i'}{f_i'}$
with ${s_P}_i'$ such that $s_i' = {s_P}_i'\pileconcat s$
\end{itemize}

\regleinfcond{infcommcc}
{P \spcontexte{C_P} \atransition{g\envoi v} P' \spcontexte{C_P'} \hspace{20pt}
Q \spcontexte{C_Q} \atransition{g\recep v} Q' \spcontexte{C_Q'}}
{P \para Q \spcontexte{C} \tautransition P' \para Q' \spcontexte{C'}}
{P\stable, Q\stable}
where $v \in \n$, $C_Q' = \contexte{s'} {q'}{\rho'}{f'}$, and
$C' = \contexte{\pileajoutp s {(s_P\para s_Q)}} {q}{\rho}{f'}$
\regleinfcond{infcommcq}
{P \spcontexte{C_P} \atransition{g\envoi v} P' \spcontexte{C_P'}\hspace{20pt}
Q \spcontexte{C_Q} \atransition{g\recep y} Q' \spcontexte{C_Q'}}
{P \para Q \spcontexte{C} \tautransition P' \para Q' \spcontexte{C'}}
{P\stable, Q\stable}
where
\begin{itemize}
\item $(x,\qubittype) \in s \cup s_Q$, $x \not\in q$, $v\in\ensemble{0,1}$
\item $C' = \contexte{\pileajoutp s {(s_P\para s_Q)}} {x.q}{\ket v\bra v\otimes\rho}{f}$
\end{itemize}
\regleinfcond{infcommqq}
{P \spcontexte{C_P} \atransition{g\envoi x} P' \spcontexte{C_P'}\hspace{20pt}
Q \spcontexte{C_Q} \atransition{g\recep y} Q' \spcontexte{C_Q'}}
{P \para Q \spcontexte{C} \tautransition P' \para Q' \spcontexte{C'}}
{P\stable, Q\stable}
where
\begin{itemize}
\item $(x,\qubittype) \in s \cup s_P$, $x\in q$
\item $(y,\qubittype) \in s \cup s_Q$, $y\not\in q$
\item$C' = \contexte{\rmpile{(\pileajoutp s {(s_P\para s_Q)})}{\ensemble x}}
{q[x\leftarrow y]}{\rho}{f}$
\end{itemize}
\regleinfcond{infcompparaprob1}
{P \spcontexte{C_P} \ptransition p P' \spcontexte{C_P'}}
{P \para Q \spcontexte{C} \ptransition{p} P'\para Q \spcontexte{C}}
{Q\stable}
\regleinf{infcompparaprob2}
{P \spcontexte{C_P} \ptransition p P' \spcontexte{C_P'} \hspace{20pt}
Q \spcontexte{C_Q} \ptransition q Q' \spcontexte{C_Q'}}
{P \para Q \spcontexte{C} \ptransition{pq} P'\para Q' \spcontexte{C}}

\end{document}